\documentclass[twocolumn]{aastex7}

\newcommand{\msol}{\mbox{$M_\odot$}}

\newcommand{\HI}{H\,{\sc i} }
\newcommand{\hi}{H\,{\sc i}} 
\newcommand {\kms}{\ifmmode{\rm km \, s^{-1}}\else{$\rm km \, s^{-1}$}\fi} 
\newcommand{\Msun} {M_{\sun}}

\shortauthors{Xu et al.}
\usepackage{textcomp, gensymb}
\usepackage{booktabs}
\usepackage{soul}
\usepackage{float}
\usepackage{comment}
\usepackage{array}
\usepackage{makecell}
\usepackage{amsmath}
\usepackage{multirow} 
\usepackage{enumitem}

\defcitealias{Xu_2023}{X23}
\defcitealias{Nakahara_2018ApJ...854..148N}{N18}

\begin{document}
\title{Gas Accretion from a Neighbouring  Galaxy Fuels the Low-luminosity AGN in NGC 4278}
\correspondingauthor{Jin-Long Xu}
\email{xujl@bao.ac.cn}

\author[0000-0002-7384-797X]{Jin-Long Xu$^{*}$}
\affiliation{National Astronomical Observatories, Chinese Academy of Sciences, Beijing 100101, People's Republic of China}
\affiliation{Guizhou Radio Astronomical Observatory, Guizhou University, Guiyang 550000, People's Republic of China}
\email{xujl@bao.ac.cn}

\author[0000-0000-0000-0000]{Nai-Ping Yu}
\affiliation{National Astronomical Observatories, Chinese Academy of Sciences, Beijing 100101, People's Republic of China}
\affiliation{Guizhou Radio Astronomical Observatory, Guizhou University, Guiyang 550000, People's Republic of China}
\email{npyu@bao.ac.cn}

\author[0000-0001-6083-956X]{Ming Zhu}
\affiliation{National Astronomical Observatories, Chinese Academy of Sciences, Beijing 100101, People's Republic of China}
\affiliation{Guizhou Radio Astronomical Observatory, Guizhou University, Guiyang 550000, People's Republic of China}
\email{mz@nao.cas.cn}

\author[0000-0002-4428-3183]{Chuan-Peng Zhang}
\affiliation{National Astronomical Observatories, Chinese Academy of Sciences, Beijing 100101, People's Republic of China}
\affiliation{Guizhou Radio Astronomical Observatory, Guizhou University, Guiyang 550000, People's Republic of China}
\email{cpzhang@bao.ac.cn}

\author[0009-0003-6680-1628]{Xiao-Lan Liu}
\affiliation{National Astronomical Observatories, Chinese Academy of Sciences, Beijing 100101, People's Republic of China}
\affiliation{Guizhou Radio Astronomical Observatory, Guizhou University, Guiyang 550000, People's Republic of China}
\email{liuxiaolan@bao.ac.cn}

\author[0000-0000-0000-0000]{Mei Ai}
\affiliation{National Astronomical Observatories, Chinese Academy of Sciences, Beijing 100101, People's Republic of China}
\affiliation{Guizhou Radio Astronomical Observatory, Guizhou University, Guiyang 550000, People's Republic of China}
\email{aimei@nao.cas.cn}

\author[0000-0002-5387-7952]{Peng Jiang}
\affiliation{National Astronomical Observatories, Chinese Academy of Sciences, Beijing 100101, People's Republic of China}
\affiliation{Guizhou Radio Astronomical Observatory, Guizhou University, Guiyang 550000, People's Republic of China}
\email{pjiang@nao.cas.cn}

\begin{abstract}
How a seemingly `dead' host galaxy provides fuel for its active galactic nuclei (AGN) remains an unresolved problem. Using the Five-hundred-meter Aperture Spherical radio Telescope (FAST), we present a new high-sensitivity atomic-hydrogen (\hi) observation toward the nearby elliptical galaxy NGC 4278 and its adjacent region. From the observation, we found that  external gas accretion from a neighbouring galaxy fuels the low-luminosity AGN in NGC 4278 through tidal interactions. The accreted gas entering NGC 4278 exhibits a  rotating gas disk.  And the accreted galaxy has been gas-poor and has an \HI to stellar mass ratio of about 0.02. Due to the process of gas accretion, it is likely that relativistic jets are generated in the AGN of NGC 4278. The emission of TeV gamma rays in NGC 4278 is likely to be associated with the newly accreted \HI gas.
\end{abstract}


\section{Introduction} \label{sec:intro}
Almost all massive galaxies (with total stellar masses $M_{\ast}$ $\geq$ 10$^{9} \Msun$) contain a supermassive black hole (SMBH) at their center \citep{2008ARA&A..46..475H}. These SMBHs evolve alongside their host galaxies, growing by accumulating gas from the interstellar medium (ISM) while also influencing or regulating star formation by affecting the cold gas content of the galaxies \citep{2013ARA&A..51..511K}. It is now widely accepted that active galactic nucleis (AGNs) are powered by the accretion of gas onto SMBHs \citep{1964ApJ...140..796S,1969Natur.223..690L,1984ARA&A..22..471R}.  Elliptical galaxies were considered gas-poor systems dominated by older stellar populations. However, they are now known to occasionally host low-luminosity AGNs (LLAGNs) \citep{1995MNRAS.277L..55F,2008ARA&A..46..475H}. A key question remains: how does a seemingly ``dead" host galaxy fuel an AGN?

Fuel for the AGN can be provided in both internal and external forms. In later-type and gas-rich galaxies, the secular processes (disk instabilities and associated inflows) dominate for the AGNs, while the major mergers, minor mergers as well as accretion of intergalactic gas may play an important role in the ANGs triggering and feeding in early-type and gas-poor galaxies \citep{2019NatAs...3...48S}. For early-type elliptical galaxies, fueling AGN through external means may be the most important approach. While the accretion of cold gas can feed the AGNs mainly in galaxy clusters \citep{2016Natur.534..218T}. The major mergers may be the most important process for luminous AGN since it require large amounts of gas to be transported to the vicinity of the SMBH in a short period of time. At lower luminous AGNs, the minor mergers may be the main approach \citep{2019NatAs...3...48S}. Hence, identifying the source of the fuel gas in the gas-poor elliptical galaxies plays an important role in how the AGN is excited within them. 

NGC 4278  is an early-type elliptical galaxy,  classified as E1-2 and located at distance of 16.4$\pm$0.2 Mpc \citep{2001ApJ...546..681T}. It  harbors an SMBH with a mass of about 3.1 $\times~10^{8}~\Msun$ \citep{2003MNRAS.340..793W,2005ApJ...625..716C}   and hosts a low-luminosity AGN  \citep{1997ApJS..112..391H,2010A&A...517A..33Y,2012ApJ...758...94P}. The two-sided symmetric radio jets on a subparsec scale has been resolved from its core region \citep{2005ApJ...622..178G}. Furthermore, the TeV gamma rays from 1LHAASO J1219+2915 is found to be spatially coincident with NGC 4278 \citep{2024ApJ...971L..45C}. Recently, \citet{2025arXiv250702326S} show that NGC 4278 has a massive molecular cloud surrounding  its nucleus from CO observation.  More notably, NGC 4278 was known to have an extended regular disc of \HI gas, with a total mass of 6.9 $\times~10^{8}~\Msun$  \citep{1981ApJ...246..708R,2006MNRAS.371..157M,2012MNRAS.422.1835S,2014MNRAS.444.3388S}. Some faint \HI tails and clouds are detected around the \HI disk of NGC 4278  \citep{2006MNRAS.371..157M,2012MNRAS.422.1835S,2014MNRAS.444.3388S}, indicating that it likely to be ongoing interactions with the surrounding environment. Therefore, NGC 4278 provides an excellent opportunity to study the origin of the AGN fueling gas.

In this paper, based on the high-sensitivity \HI observations using the FAST, we first show that the external gas accretion from a neighbouring  galaxy fuels the low-luminosity AGN in the nearby elliptical galaxy NGC 4278 by analyzing data.

\section{Observation and data processing}
\subsection{Observations and data processing of FAST}
To detect the diffuse cold gas around the periphery of  NGC 4278, we have performed the \HI (1420.4058 MHz) observations towards it using the Five-hundred-meter Aperture Spherical radio Telescope (FAST) \citep{2019SCPMA..6259502J,2020RAA....20...64J} during the May 2024. The new \HI observations used the Multi-beam on-the-fly (MultibeamOTF) mode. This mode is designed to map the sky with 19 beams simultaneously. It formally works in the frequency range from 1050 MHz to 1450 MHz. The half-power beam width (HPBW) is $\sim$2.9$^{\prime}$ at 1.4 GHz for each beam. The scan velocity and an integration time for per spectrum are set as 15$^{\prime\prime}$ s$^{-1}$ and 1 second, respectively.  We used the digital Spec (W) backend, which has a bandwidth of 500 MHz and 64k channels, resulting in a velocity resolution of about 1.6 \kms. For intensity calibration to antenna temperature ($T_{\rm A}$), noise signal with amplitude of 10 K was injected in a period of 64 seconds. The system temperature was around 22 K for the \HI survey. The pointing accuracy of the telescope was better than 10$^{\prime\prime}$. 

The data reduction was performed with the PYTHON packages. Finally, we made the calibrated data into the standard cube data, with a pixel of $1.0^{\prime}\times1.0^{\prime}$. The detailed data reduction is similar to \citet{2021ApJ...922...53X}. A gain $T_{\rm A}/\it S_{v}$ has been measured to be about 16 K Jy$^{-1}$. While a measured relevant main beam gain $T_{\rm B}/\it S_{v}$ is about 21 K Jy$^{-1}$ at 1.4 GHz for each beam, where $T_{\rm B}$ is the brightness temperature. To creat a highly sensitivity \HI image, the velocity resolution of the FAST data is smoothed to 6.4 \kms. In the observed region, we measured that the root-mean-square (RMS) noise is  0.45 mJy beam$^{-1}$.

\subsection{Other archive data}
To verify some key gas components, we also use high-resolution data from the Westerbork Synthesis Radio Telescope (WSRT) observations \citep{2006MNRAS.371..157M}. The \HI cube data has an angular resolution of $\sim43^{\prime\prime}\times33^{\prime\prime}$ and a velocity resolution of $\sim$12.4 \kms. The optical image of the Dark Energy Spectroscopic Instrument (DESI) is derived from the Legacy Surveys Sky Viewer websites\footnote{http://viewer.legacysurvey.org}.

\begin{figure}
\centering
\includegraphics[width=0.5\textwidth]{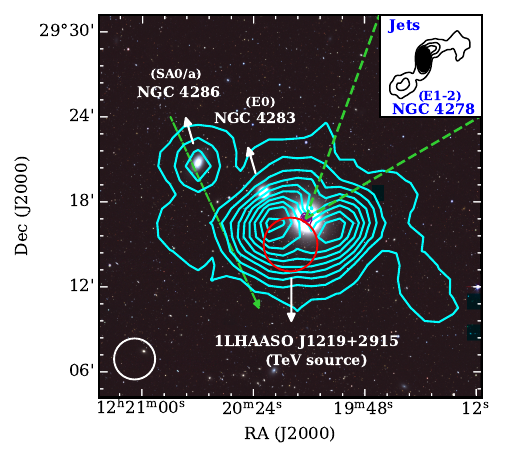}
\setlength{\abovecaptionskip}{-8pt}
\caption{Large-scale overview of NGC 4278.  \HI column-density map of NGC 4278 from the FAST observation shown in cyan contours overlaid on the DESI-RGB ($g, r, z$) image in color scale. The cyan contours begin at 5.4$\times$10$^{17}$ cm$^{-2}$ (3$\sigma$) in steps of 3.8$\times$10$^{18}$ cm$^{-2}$. The red circle represents TeV source 1LHAASO J1219+2915, while the purple contours indicate the X-ray emission. The radio jets of NGC 4278 are shown in the upper-right corner. The green arrow indicates a cuting direction and position of the position-velocity diagram shown in Fig. 4. The FAST beam in a white circle is shown in the bottom-left corner.}
\label{fig:NGC4278-HI-optical}
\end{figure}

\section{Results}
\label{sect:results}

Figure \ref{fig:NGC4278-HI-optical} shows the large-scale overview of NGC 4278 from the different bands. 
The three-dimensional source finder SoFiA2  was applied to the FAST data cube to identify the emission regions of NGC 4278 \citep{2021MNRAS.506.3962W}. During this process, we set a primary threshold to identify emission peaks at 3$\sigma$ RMS.  The column density $N(x,y)$ in each pixel of the image can be estimated as $N(x,y)= 1.82\times10^{18}\int T_{\rm B}dv$,  where $dv$ is the velocity width in \kms. In Fig. \ref{fig:NGC4278-HI-optical}, the \HI column-density map from the FAST data in cyan contours is overlaid on the DESI-RGB image in colorscale. Using an RMS noise of 0.45 mJy beam$^{-1}$, it corresponds to a $1\sigma$ \HI\ column density of $\rm 1.8\times10^{17}\,cm^{-2}$ per 20 km s$^{-1}$. With this high detection sensitivity, we can see that there is a large \HI gas complex, which overlaps with the optical emission from galaxies NGC 4278, NGC 4283, and NGC 4286 projected on the sky. The basic parameters of the three galaxies are  available at NASA/IPAC Extragalactic Database\footnote{http://ned.ipac.caltech.edu/} and listed in Table \ref{tab:prop}. The main structure is associated with NGC 4278, but its peaks seem to deviate from its optical center, which is associated with previous \HI observations \citep{2006MNRAS.371..157M,2012MNRAS.422.1835S,2014MNRAS.444.3388S}. It's very interesting that one of the peaks is very close to the TeV gamma rays from 1LHAASO J1219+2915 \citep{2024ApJ...971L..45C}, which is shown in  a red circle. Moreover, the northeast structure associated with SA0/a NGC 4286 appears to connect the \HI structure of NGC 4278 projected on the sky, while the southwest gas show a tail structure, which is associated with that from the WSRT observations  \citep{2006MNRAS.371..157M,2012MNRAS.422.1835S,2014MNRAS.444.3388S}. Previously,  some clouds are detected around the \HI disk of NGC 4278. Here we did not detect them. There are two possibilities, one is that these clouds are some noise, and the other is that our  used telescopes cannot distinguish them because these clouds are relatively small. Furthermore, NGC 4283 is also an elliptical galaxy, whose system velocity is 1056.0 \kms. On the one hand, we did not detect any corresponding \HI gas emission at the system velocty. On the other hand, it was significantly higher than the overall velocity range of this \HI complex. Hence, it suggests that NGC 4283 is a gas-poor background galaxy.

\begin{table}
\footnotesize
\centering
\caption{\small Measured and derived properties of the known galaxies. We listed: equatorial coordinates (RA, Dec.);  system velocity ($V_{\rm sys}$); line widths  at 20\% of the peak flux ($W_{20}$) and at 50\% of the peak flux ($W_{50}$); total \HI flux ($S_{\rm tot}$); apparent magnitude ($m_{g}$, $m_{r}$), which has been corrected for extinction; \HI gas mass ($M_{\rm HI}$); steller mass ($M_{\rm \star}$);}
\label{tab:prop}
\setlength{\tabcolsep}{4pt}
\begin{tabular}{lcccc}
\noalign{\vspace{1pt}}\hline\hline\noalign{\vspace{5pt}}
Name & NGC 4278 & NGC 4283  & NGC 4286 \\
\noalign{\vspace{1pt}}\hline\noalign{\vspace{5pt}}
Type &  E1-2    & E0     &  SA0/a \\
RA   & 12$^{\rm h}$20$^{\rm m}$06.8$^{\rm s}$ & 12$^{\rm h}$20$^{\rm m}$20.8$^{\rm s}$ & 12$^{\rm h}$20$^{\rm m}$42.1$^{\rm s}$ \\
Dec. & 29$^{\rm \circ}$16$^{\rm \prime}$50.7$^{\rm \prime\prime}$ & 29$^{\rm \circ}$18$^{\rm \prime}$39.4$^{\rm \prime\prime}$ & 29$^{\rm \circ}$20$^{\rm \prime}$45.2$^{\rm \prime\prime}$ \\
$V_{\rm sys}$[\kms] & 629.7$^{+4.0}_{-4.0}$ & 1056.0$^{+5.0}_{-5.0}$& 637.0$^{+6.0}_{-6.0}$ \\
$W_{\rm 50}$ [$\kms$] & 360.5$_{-1.2}^{+1.2}$ & --- & 109.6$_{-2.5}^{+2.5}$  \\
$W_{\rm 20}$ [$\kms$] & 422.8$_{-1.4}^{+1.4}$ & --- & 123.1$_{-2.9}^{+2.9}$\\
$S_{\rm tot}$ [$\rm Jy~\kms$] & 11.5$_{-0.4}^{+0.4}$ &--- & 0.7$_{-0.1}^{+0.1}$\\
$m_{g}$[mag]  & 10.45 & 12.36 & 13.23 \\
$m_{r}$[mag]  & 9.72 & 11.65 & 12.68\\
log($M_{\rm HI}$)[$\msol$]  &8.9$^{+0.1}_{-0.1}$& --- & 7.6$^{+0.1}_{-0.1}$ \\
log($M_{\rm \star}$)[$\msol$] &10.7$^{+0.1}_{-0.1}$ & 9.9$^{+0.1}_{-0.1}$&  9.3$^{+0.1}_{-0.1}$\\
\noalign{\vspace{5pt}}\hline\noalign{\vspace{1pt}}
\end{tabular}
\end{table}

\begin{figure}
\centering
\includegraphics[width=0.45\textwidth]{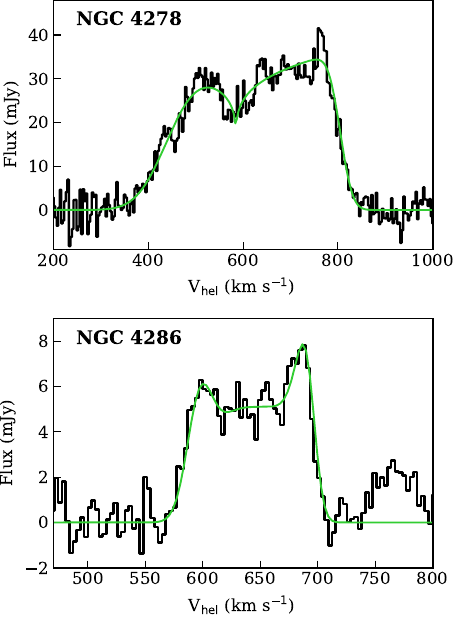}
\setlength{\abovecaptionskip}{0pt}
\caption{Global \HI profiles of NGC 4278 and NGC 4286 shown in a black lines. The green lines indicate the BusyFit fitting result. The values of the errorbars are 0.4 Jy for NGC 4286, while 1.0 Jy for NGC 4278. }
\label{fig:NGC4278-HI-spectrum}
\end{figure}

Both NGC 4278 and NGC 4286 contain the \HI gas. In Fig. \ref{fig:NGC4278-HI-spectrum}, we show the global \HI profile of NGC 4278 and NGC 4286. Both the \HI profiles seem to exhibit a double-peak shape, which is generally thought to be caused by the flat rotation curve of a disk galaxy. NGC 4286 is a SA0/a galaxy. It is normal for it to contain a rotating gas structure. We performed a BusyFit fitting to the profile to determine the total flux ($S_{\rm tot}$), line widths at 50\% of the peak flux ($W_{50}$) and at 20\% of the peak flux ($W_{20}$). The busy function is an analytic function for describing the integrated \HI spectral profile of galaxies \citep{2014MNRAS.438.1176W}. The parameters obtained by the fitting are listed  in Table \ref{tab:prop}. 

\begin{figure*}
\centering
\includegraphics[width=0.95\textwidth]{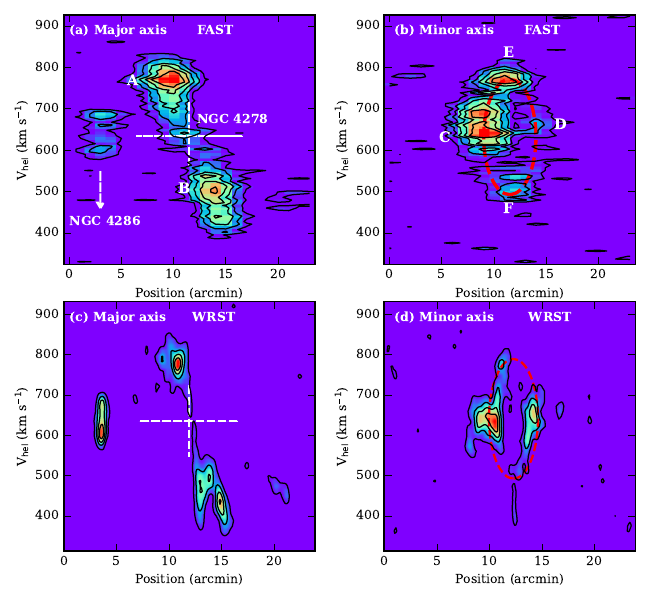}
\setlength{\abovecaptionskip}{-11pt}
\caption{Position-velocity diagrams of NGC 4278 in color scale overlaid with the black contours. {\bf a}, the direction of cutting for the slice from the FAST data is through the center of the galaxy along the major axis (PA=65$^{\circ}$). The black contours begin at 3$\sigma$ ({\bf 1.4 mJy beam$^{-1}$}) in steps of 3$\sigma$. {\bf b}, the direction of cutting for the slice from the FAST data  is through the center of the galaxy along the minor axis (PA=140$^{\circ}$).  {\bf c}, the slice from the WRST data along the major axis. The black contours begin at 3$\sigma$ (0.3 mJy beam$^{-1}$) in steps of 4$\sigma$. {\bf d}, the slice from the WRST data along the major axis. The large white pluses in panels a and c mark the centre and systemic velocity of NGC 4278, while the red circles in panels b and d indicates a fitting shell.}
\label{fig:NGC4278-Moment-pv}
\end{figure*}

In order to  further explore the dynamic structure of NGC 4278, we made the position-velocity (PV) diagrams, as shown in Fig. \ref{fig:NGC4278-Moment-pv}. The extracted PV diagrams pass through the center of NGC 4278 with a length of 24.0$^{\prime}$ along its major axis (PA = 65$^{\circ}$) and minor axis (PA = 140$^{\circ}$). The position angle (PA) is obtained from the WSRT observation \citep{1981ApJ...246..708R}. We use different letters to indicate the corresponding gas components. From the PV diagrams along the major axis in Figures. \ref{fig:NGC4278-Moment-pv}a and \ref{fig:NGC4278-Moment-pv}c from the FAST and WRST data, we see that NGC 4278 seems to have two velocity components.  Components A and B form an S-shaped rotating structure, which is just associated with the double-peak shape in its global \HI profile. The S-like structure is a typical characteristics of disk galaxies \citep{2020Natur.581..269N}. In the PV diagram along the minor axis (Fig. \ref{fig:NGC4278-Moment-pv}b), we see that there seem to be four components here, designed as C, D, E, and F, which exactly form a ring, as indicated by a red dashed circle.  These components are also shown in the Fig. \ref{fig:NGC4278-Moment-pv}d from the WRST high-resolution data. For a disk galaxy \citep{2020Natur.584..201R}, we usually see components C and D in the PV diagram along the minor axis of its disk, without components E and F. In comparing components A and B with components E and F, as shown in Figures 3a and 3b, we found that their velocities are consistent with the system velocity of NGC 4278. In the PV disgrams, based on a S-like rotation structure along the main axis and  a ring structure along the minor axis,  we suggest that the \HI gas surrounding NGC 4278 is a  weird rotating gas disk.

Gas content of galaxies can be used to explore their evolutionary history.
The gas in galaxies mainly consists of \HI and helium.  Assuming the same helium-to-\HI ratio as that derived from the Big Bang nucleosynthesis, a factor of 1.33 is included to account for the contribution of helium. The total gas mass is determined with $M_{\rm gas} = 1.33\times M_{\rm HI}$. Here the \HI gas mass of galaxies can be estimated as  $M_\mathrm{HI}=2.36\times10^{5}D^{2}{\int}S_{v}dv$,  where  ${\int}S_{v}dv$ is the integrated {\HI} flux in Jy km s$^{-1}$, and $D$ is the adopted distance in Mpc to each galaxy. The total flux $S_{\rm tot}$ of NGC 4278 and NGC 4286 are 11.5$\pm$0.4 Jy km s$^{-1}$ 0.7$\pm$0.1 Jy km s$^{-1}$ obtained from their global \HI profile, respectively. The distance to NGC 4278 is 16.4$\pm$0.2 Mpc \citep{2001ApJ...546..681T}. Since NGC 4278 and NGC 4286 are spatially connected, we assume that their distances are the same.
Finally, the obtained $M_{\rm gas}$  are 9.7$\times10^{8} \Msun$ and 5.8$\times10^{7} \Msun$ for NGC 4278 and NGC 4286, respectively. In Table \ref{tab:prop}, the uncertainties of $W_{50}$, $W_{20}$, and $S_{\rm tot}$ are from the  Busy-Function fitting.  For $M_{\rm HI}$, the errors are considered taking into account distance errors and $S_{\rm tot}$ errors.  Moreover, in order to calculate the stellar masses  ($M_*$) in NGC 4278 and NGC 4286, we use the DESI r-band and g-band archive images. The stellar masses are estimated using the mass-to-light ratio equation  $\log(M_*/L_g)$=-0.601+1.294($g-r$), where $L_g$ is the $g$ band luminosity derived from the absolute magnitude \citep{2016AJ....152..177H}. We derived that $M_*$ are  5.0$\times10^{10} \Msun$ and 1.9$\times10^{9} \Msun$ for NGC 4278 and NGC 4286, respectively. Here the errors of $M_*$ is only considering the error caused by distance. We found that their \HI gas masses are nearly two orders of magnitude less than the stellar mass, implying that these are two galaxies that are extremely gas poor.

\begin{figure*}
\centering
\includegraphics[width=0.95\textwidth]{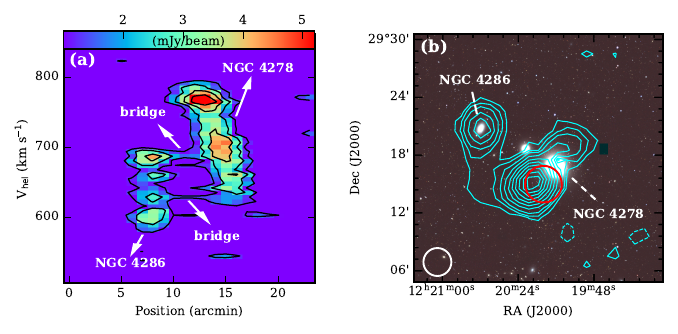}
\setlength{\abovecaptionskip}{-10pt}
\caption{{\bf a}, the velocity-position diagram in color scale overlaid with the black contours. The direction of cutting for the slice is along the direction of the accretion, as shown in Fig 1. The black contours begin at 3$\sigma$ (1.4 mJy beam$^{-1}$) in steps of 3$\sigma$.  {\bf b},  \HI column-density map in cyan and blue contours overlaid on the DESI-RGB image in color scale. The cyan solid contours from the FAST data begin at 1.1$\times$10$^{18}$ cm$^{-2}$ (3$\sigma$) in steps of 1.1$\times$10$^{18}$ cm$^{-2}$, while the cyan dashed contours begin -3$\sigma$. The integrated velocity for the cyan contours ranges from 620.0 $\kms$ to 695.0 $\kms$. The red circle represent Tev source 1LHAASO J1219+2915, whose size represents the positional error of this source. The FAST beam in a white circle is shown in the bottom-left corner.}
\label{fig:small-PV-NH2}
\end{figure*}

\begin{figure}
\centering
\includegraphics[width=0.5\textwidth]{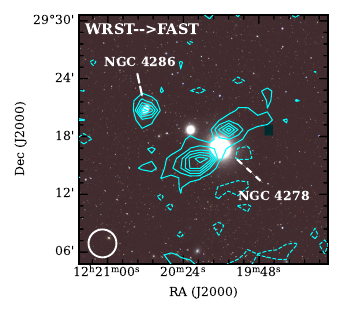}
\setlength{\abovecaptionskip}{-10pt}
\caption{The WRST \HI column-density map in cyan contours overlaid on the DESI-RGB image in color scale. The WRST cube data is  convolved to the same beam as FAST.  The cyan contours begin at 3.0$\times$10$^{18}$ cm$^{-2}$ (3$\sigma$)  in steps of 6.0$\times$10$^{18}$ cm$^{-2}$,  while the cyan dashed contours begin -3$\sigma$. The integrated velocity for the cyan contours ranges from 620.0 $\kms$ to 695.0 $\kms$. The FAST beam in a white circle is shown in the bottom-left corner.}
\label{fig:FF-PV-NH2}
\end{figure}

\section{Discussion and Conclusion}
\label{sect:discussion}
For early-type elliptical galaxies, fueling of AGN  is mainly achieved through the major mergers, minor mergers, and accretion of intergalactic gas. The accretion of cold gas can feed the AGNs mainly from the hot intergalactic medium in galaxy clusters \citep{2016Natur.534..218T,2019NatAs...3...48S}. The X-ray emission is only detected in the center of NGC 4278 \citep{2019ApJS..241...36K}, as shown in Fig \ref{fig:NGC4278-HI-optical}. Therefore, the relatively isolated elliptical galaxy NGC 4278 cannot have formed by the accretion of intergalactic gas. Furthermore, the AGNs feeding through the major mergers are generally luminous AGN, and the merged gas is almost gas-poor and no regular shape around the host galaxies \citep{2016Natur.534..218T}. 
NGC 4278 has a low-luminosity AGN \citep{1997ApJS..112..391H,2010A&A...517A..33Y,2012ApJ...758...94P} and a rotating gas disk. Therefore, NGC 4278 likely did not originate through a major merger. Additionally, given the regular optical shape of NGC 4278 and the absence of detected stripped stellar streams in its vicinity, we conclude that minor mergers are also not a source of its relatively low-luminosity AGN.   Therefore, there may be alternative mechanisms at play that contribute to the feeding of the AGN in NGC 4278. 

Both WSRT and FAST observations detected a faint \HI tail around the \HI disk of NGC 4278  \citep{2006MNRAS.371..157M,2012MNRAS.422.1835S,2014MNRAS.444.3388S}, implying that it likely to be ongoing interactions with the surrounding environment.
From the \HI observations conducted using the FAST, NGC 4278 and NGC 4286 are found to be connected on the sky, as illustrated in Fig. \ref{fig:NGC4278-HI-optical}. To analyze the kinematics of these two galaxies in velocity space, we created a PV diagram through their centers with a length of 21$^{\prime}$. The cutting direction is indicated by a green dashed arrow in Fig. \ref{fig:NGC4278-HI-optical}. The PV diagram in Fig. \ref{fig:small-PV-NH2}a shows that NGC 4278 and NGC 4286 are not only connected spatially but also in terms of velocity. This connection implies the presence of an \HI gaseous bridge between the two galaxies. The velocity range of the connecting bridge is between 620 \kms and 695 \kms. Using this velocity range, we created a column-density map from the FAST data, as shown in Fig. \ref{fig:small-PV-NH2}b, and obtained a $1\sigma$ \HI\ column density of $\rm 3.6\times10^{17}\,cm^{-2}$. The column-density map in the cyan contours shows a curved bridge, not directly connecting two galaxies together, which is larger than the size of two FAST beams. In addition, we convolved the WSRT cube data to the same beam and pixel as the FAST data, and created a column-density map, as shown in  Figure \ref{fig:FF-PV-NH2}. The \HI image in Fig. \ref{fig:FF-PV-NH2} was made integrating over the same velocity range used to make Fig. \ref{fig:small-PV-NH2}b. We see that these are two small gas clouds between NGC 4278 and NGC 4286, whose positions are consistent with the curved bridge detected by FAST.  From the above signs, we conclude that this bridge detected by FAST is a true gas bridge connected  NGC 4278 with NGC 4286,  not caused by FAST beam smearing. The detected bridge suggests that NGC 4278 and NGC 4286 are interacting, which is considered a mechanism for massive galaxies to replenish their gas reserves \citep{2008A&ARv..15..189S}. Additionally, the peak position of this part \HI gas aligns with the TeV gamma rays observed from 1LHAASO J1219+2915 within the uncertain range of the rays position detection, as shown in a red circle in Fig. \ref{fig:small-PV-NH2}b.  The TeV emissions is likely to be associated with the newly accreted \HI gas.

Adjacent to NGC 4278 on the sky, there is an elliptical galaxy NGC 4283, whose system velocity is larger than 426.3 \kms~  that of NGC 4278. Assuming the same distance as NGC 4278, we obtain a minimum stellar mass of  7.9$\times10^{9} \Msun$ for NGC 4283. In these two massive elliptical galaxies with similar environments and stellar mass, they should also have similar formation and evolution pathways. We did not detect any corresponding \HI gas emission in NGC 4283, whereas NGC 4278 currently possesses an abundant gas supply. This suggests that the gas in NGC 4278 is not merely residual gas following its formation and evolution. Moreover, the \HI profile of NGC 4286 displays a double-peak shape typical of disk galaxies. Previous studies characterized the morphology of NGC 4286 as belonging to the SA0/a classification, indicating it is a transitional type between a lenticular galaxy (SA0) and a spiral galaxy (Sa) \citep{2015ApJS..217...27A}, and  has already started to gradually lose gas before.   NGC 4286 has a stellar mass of 1.9$\times10^{9} \Msun$. The \HI to stellar mass ratio for this galaxy is only about 2\%, which suggests a significant deficiency in \HI gas. This deficient \HI indicates that a large portion of gas has been stripped from NGC 4286 and is being accreted by NGC 4278 through tidal interactions.

The bipolar symmetric radio jets have been resolved from the AGN of NGC 4278 \citep{2005ApJ...622..178G}. They found low apparent velocities ($\lesssim$0.2$c$) for the jet components and estimate the epoch of their ejection as 10-100 yr prior to their observations.  Previous studies suggested that AGNs are powered by the accretion of gas onto SMBHs \citep{1964ApJ...140..796S,1969Natur.223..690L,1984ARA&A..22..471R}. It suggests that recent gas accretion has provided fuel for the AGN of NGC 4278. While we found that NGC 4278 is accreting gas from NGC 4286 through tidal interactions.  Due to the resolution limitations of our data (FAST and WRST), we are currently unable to decipher how the accreted gas is transported from the outer disk to the inner disk and smaller scale nuclear region.  However, for NGC 4278, the present observations indicate a position angle of 65$^{\circ}$ for the kinematic major axis of its outer gas, 45$^{\circ}$ away from the optical major axis \citep{1981ApJ...246..708R}. Numerical simulations have suggested that misaligned structures between gas disk and stellar disk may promote the inflow of gas to the nucleus of the galaxy and the accretion of gas by the central supermassive black hole \citep{2023NatAs...7..463R}. In addition, on galactic scales, numerical simulations also have shown that tidal torques can lead to rapid gas inflow into the central kpc \citep{1991ApJ...370L..65B,1996ApJ...471..115B}. Hence, we propose that the accreted gas from NGC 4286 is flowing toward the nucleus region of NGC 4278, which fuels the low-luminosity AGN and triggers the bipolar jet of the AGN.

\begin{acknowledgments}
We thank the referee for insightful comments that improved the clarity of this manuscript.
We acknowledge the support of the National Key R$\&$D Program of China No. 2022YFA1602901. This work is also supported by the National Natural Science Foundation of China (Grant Nos. 12373001, 12225303, 12421003), the Chinese Academy of Sciences Project for Young Scientists in Basic Research, grant no. YSBR-063, the Guizhou Provincial Science and Technology Projects (
Supported by the Guizhou Provincial Science and Technology Projects (No.QKHFQ[2023]003, No.QKHFQ[2024]001, No.QKHPTRC-ZDSYS[2023]003, No.QKHJC-ZK[2025]MS015).  
\end{acknowledgments}

\bibliography{references} 
\bibliographystyle{aasjournal}
\end{document}